\documentclass{webofc}
\usepackage[varg]{txfonts}   
\usepackage{fancyvrb}
\usepackage{xcolor}

\definecolor{MyGreen}{RGB}{56,118,29}
\definecolor{MyRed}{RGB}{153,0,0}
\definecolor{MyPurple}{RGB}{53,28,117}

\newcommand{\edmWaitingTaskWithArenaHolder}{\texttt{edm::WaitingTaskWithArenaHolder}}
\newcommand{\cmsProductT}{\texttt{cms::cuda::Product<T>}}
\newcommand{\cmsScopedContext}{\texttt{cms::cuda::ScopedContext}}
\newcommand{\WaitingTaskWithArenaHolder}{\texttt{WaitingTaskWithArenaHolder}}
\newcommand{\ProductT}{\texttt{Product<T>}}
\newcommand{\ScopedContext}{\texttt{ScopedContext}}

\begin{document}
\title{Bringing heterogeneity to the CMS software framework}
%
%


\author{\firstname{Andrea} \lastname{Bocci}\inst{1} \and
        \firstname{David} \lastname{Dagenhart}\inst{2} \and
        \firstname{Vincenzo} \lastname{Innocente}\inst{1} \and
        \firstname{Christopher} \lastname{Jones}\inst{2} \and
        \firstname{Matti} \lastname{Kortelainen}\inst{2} \fnsep\thanks{\email{matti@fnal.gov}} \and
        \firstname{Felice} \lastname{Pantaleo}\inst{1} \and
        \firstname{Marco} \lastname{Rovere}\inst{1}
}

\institute{CERN, Geneva, Switzerland
\and
           Fermi National Accelerator Laboratory, Batavia, IL, USA
          }

\abstract{%
The advent of computing resources with co-processors, for example Graphics Processing Units (GPU) or Field-Programmable Gate Arrays (FPGA), for use cases like the CMS High-Level Trigger (HLT) or data processing at leadership-class supercomputers imposes challenges for the current data processing frameworks. These challenges include developing a model for algorithms to offload their computations on the co-processors as well as keeping the traditional CPU busy doing other work. The CMS data processing framework, CMSSW, implements multithreading using the Intel Threading Building Blocks (TBB) library, that utilizes tasks as concurrent units of work. In this paper we will discuss a generic mechanism to interact effectively with non-CPU resources that has been implemented in CMSSW. In addition, configuring such a heterogeneous system is challenging. In CMSSW an application is configured with a configuration file written in the Python language. The algorithm types are part of the configuration. The challenge therefore is to unify the CPU and co-processor settings while allowing their implementations to be separate. We will explain how we solved these challenges while minimizing the necessary changes to the CMSSW framework. We will also discuss on a concrete example how algorithms would offload work to NVIDIA GPUs using directly the CUDA API. 
}
\maketitle
\section{Introduction}
\label{intro}

Co-processors or computing accelerators like graphics processing units
(GPU) or field-programmable gate arrays (FPGA) are becoming more and
more popular to keep the cost and power consumption of computing
centers under control. For example, GPUs are used in many leading
supercomputers, are being used in a trigger farm by
ALICE~\cite{alice,alicegpu12,alicegpu15,alicegpu17,alicegpu20}, and are being considered in trigger farms for the
Run 3 data taking of the Large Hadron Collider~\cite{lhc}
in CMS~\cite{cms,bocci:2020} and LHCb~\cite{lhcb,allen}. The CMS data processing
framework
(CMSSW)~\cite{jones:2014,jones:2015,jones:2017,jones:2018,dagenhart:2020}
implements multi-threading using the Intel Threading Building Blocks
(TBB)~\cite{tbb} library utilizing tasks as units of concurrent work.
While in principle non-CPU resources could be interacted with in the TBB
tasks directly in a straightforward way, the non-CPU APIs typically
imply blocking the calling thread. Such blocking would lead to
under-utilizing the CPU.

In this paper we describe generic mechanisms to interact with non-CPU
resources effectively from the TBB tasks (Section~\ref{externalWork}),
and to configure CPU and non-CPU algorithms in a unified way that
works well together with the rest of the CMS computing infrastructure
(Section~\ref{switchProducer}). As a first step to gain experience, we
have explored various ways for how algorithms could offload work to
NVIDIA GPUs with CUDA~\cite{cuda}. Section~\ref{cudaTools} describes a pattern
that we have found most effective so far, and has also the least
impact on the rest of the CMSSW framework.

\section{Concurrent CPU and non-CPU processing}
\label{externalWork}

When computations are offloaded to non-CPU resources, the CPU program
needs to eventually know when the offloaded work is finished. The
simplest way to perform this synchronization is to introduce a
\textit{blocking wait} on the CPU thread\footnote{E.g. in CUDA
  \texttt{cudaDeviceSynchronize()}, \texttt{cudaStreamSynchronize()},
  \texttt{cudaEventSynchronize()}.}, i.e. the CPU thread waits for the
completion of the offloaded work. The CPU thread can wait either by
busy waiting or sleeping. The downside of the former approach is that
the CPU core is unable to do other work, implying that such waits
should be short at best, while the downside of the latter approach is
that the latency from the work completion to the CPU thread resuming
work is longer than in the former approach. CMS data processing
applications typically have some work that could be done
concurrently with the offloaded work, and therefore the busy waiting
would clearly lead to wasting CPU resources.

In the case of CMS applications, the thread-sleeping approach also has a
subtle downside. The number of available CPU cores is decided
externally to the application, and may be less than the total number
of logical CPU cores of the compute node. In addition, we can not
assume that the compute node would enforce the limit on the number of CPU cores the CMS
application is allowed to use, instead the CMS application should act
as a good citizen and keep at most the allowed number of CPU cores
busy on the average. With CPU-only work good CPU utilization can be
achieved simply by initializing the TBB thread pool to use the same
number of threads as the number of cores, and letting the TBB task
scheduler keep the threads busy. In this way all the CPU cores are
kept utilized as long as there are enough tasks to fill the threads,
without a risk of using additional CPU cores. Offloading computational
work and synchronizing the CPU thread by sleeping to wait for the
offloaded work to finish would lead to under-utilization of the CPU
cores. In principle the application could be configured to use more
threads than allowed CPU cores, but then the ratio of threads to cores
would become a tunable parameter that would depend for example on the
exact application type, the CPU performance, the offloaded-to-resource
performance, and the data being processed. In order to avoid
introducing such an additional tunable parameter, we chose to develop
a generic mechanism that allows the CPU thread to run other TBB tasks
while the offloaded work is being run elsewhere


The basic idea of the \textit{External Worker} concept is to replace
the blocking waits with a callback-style
solution
. Traditionally the algorithms scheduled by the CMSSW framework
(called \textit{modules}) have one function that is called by the
framework for each event. The exact function name depends on the
module type\footnote{Analyzer modules are only allowed read event data
  products and have a member function \texttt{analyze()}, producer
  modules can also insert new data products and have a member function
  \texttt{produce()}, and filter modules can also decide whether a
  trigger path should continue or stop execution and have a member
  function \texttt{filter()} }, for the simplicity in the following
only the producer module case is described. The concept itself,
however, is general and works similarly with filter and analyzer
modules as well. It could be further noted that the External Worker
concept resembles the \texttt{async\_node} in the TBB Flow Graph
library~\cite{tbb}.

The traditional \texttt{produce()} member function is split into two
stages: \textit{acquire} and \textit{produce}. First, the framework
calls an \texttt{acquire()} member function, that can only read event
data products, and should launch the offloaded work. The
\texttt{acquire()} function is given a reference-counted holder object
(\edmWaitingTaskWithArenaHolder) that holds the TBB task that will make the
framework call the \texttt{produce()} function. The holder object
is intended to be notified upon completion of the offloaded work.
Internally the holder decreases the reference count, and once the
count reaches zero, the contained TBB task is enqueued to the task arena
the holder also holds a pointer to. Thanks to the explicit use of the task arena
the holder can be given to non-TBB threads to be signaled. The holder
is also capable of delivering exceptions. See Section~\ref{cudaTools:async}
on how this mechanism can be used with CUDA.

\section{Unified configuration for CPU and non-CPU algorithms}
\label{switchProducer}

CMS uses a hash of the application configuration to segregate data
from different workflows. The simplest approach to configure jobs using
GPUs would be to create a configuration different from a CPU-only job.
In this approach, however, the data from a single dataset processed
with CPU-only and with GPU resources would have different hashes, and
therefore would be treated as different datasets. Such a feature would
significantly restrict the flexibility of the CMS data
processing workflow management system, which consists of a global pool
of jobs that can, in principle, run at any site. To preserve the
flexibility of processing parts of a dataset on any architecture, the
configuration hash must be the same for all architectures.

We wanted to be able to keep CPU and non-CPU algorithms separate to
enable an evolutionary migration path. For example, in order to
introduce non-CPU algorithms, the current, working and validated, CPU
algorithms can be left untouched. In addition, the natural work
division may differ for different hardware architectures. It could
also happen that some non-CPU architectures are in conflict in a way
that prevents dynamically loading their libraries into the same
application. On the other hand, we do not want to preclude having CPU
and non-CPU algorithm in the same module either.

The CMSSW framework already tracks the input data of each module event
by event. We decided to use the same provenance tracking mechanism to
store also information about the choice of technology. This
information enables us to inspect afterwards the architecture on which a
given event was processed.

\begin{figure}[tbh]
  \hfill
  \begin{minipage}{0.45\textwidth}
\begin{Verbatim}[fontsize=\small,commandchars=\\\{\}]
clusters = Producer("ClusterProducer",
    input = "raw"
)
\textcolor{MyGreen}{\textbf{hits}} = \textcolor{MyGreen}{\textbf{SwitchProducer}}(
    \textcolor{MyRed}{\textbf{cpu}} = Producer("\textcolor{MyRed}{\textbf{HitProducer}}",
        input = "clusters"),
    \textcolor{MyPurple}{\textbf{gpu}} = Producer("\textcolor{MyPurple}{\textbf{HitProducerGPU}}",
        input = "raw")
)
seeds = Producer("SeedProducer",
    input = "hits"
)
\end{Verbatim}
  \end{minipage}
  \hfill
  \begin{minipage}{0.45\textwidth}
    \includegraphics[width=3.5cm,clip]{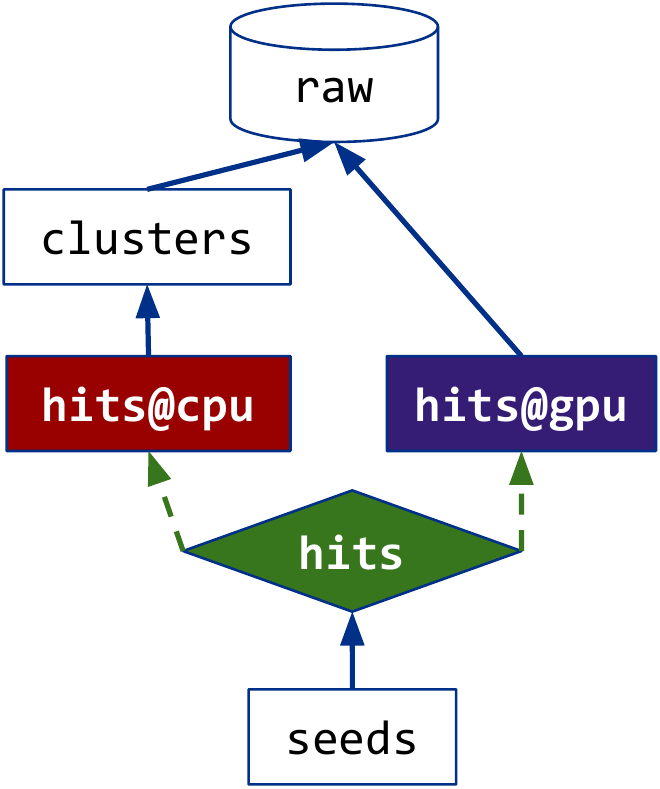}
  \end{minipage}
  \hfill
  \caption{A configuration fragment showing an example of how the
    SwitchProducer would look like (left), and a data dependence graph
    corresponding the configuration (right). On the CPU case, the
    \texttt{HitProducer} depends on a data product \texttt{clusters},
    whereas on the GPU case, the \texttt{HitProducerGPU} takes
    directly \texttt{raw} as an input. The \texttt{SwitchProducer}
    decides at runtime on the worker node which of the two producers
    should be used.%
  }
\label{fig:switchProducer}
\end{figure}

Based on the aforementioned goals, we developed the
\textit{SwitchProducer} concept in the CMSSW configuration, depicted
in Figure~\ref{fig:switchProducer}. The SwitchProducer allows
specifying multiple modules that are associated to the same module
label\footnote{In CMSSW each module must have a label that is unique
  within a process. The event data products produced by the module are
  associated to the module label.}. The modules for different cases
can be either totally different modules, or differently configured
instances of the same module. Thus all possibilities are specified in
the part of the configuration that affects the hash computation. The
mechanism makes the choice between the cases at runtime on the worker
node based on available technologies. The mechanism relies on the
CMSSW's module scheduling logic of consumer modules dictating which
producer modules are run. For example in the case of
Figure~\ref{fig:switchProducer}, if the worker node has a GPU, only
the \texttt{hits@gpu} module is run to produce the input for
\texttt{seeds} module. If, on the other hand, the worker node does not
have a GPU, both \texttt{hits@cpu} and \texttt{clusters} modules are
run.

It should be noted that the SwitchProducer requires that the producer
modules of all the cases produce exactly the same data product types
(\texttt{hits@cpu} and \texttt{hits@gpu} in in
Figure~\ref{fig:switchProducer}). This constraint ensures that the
choice by the SwitchProducer is transparent to all consumer modules
(e.g. \texttt{seeds} in Figure~\ref{fig:switchProducer}).

\section{Pattern to interact with CUDA runtime}
\label{cudaTools}

Based on the external worker (Section~\ref{externalWork}) and
SwitchProducer (Section~\ref{switchProducer}) concepts we developed
tools and a pattern to interact with the CUDA runtime from CMSSW
modules. The pattern is described as follows. We wanted the CPU to be
able to do other work while the GPU is running an algorithm. This
asynchronous execution is described in Section~\ref{cudaTools:async}.
We wanted to minimize data movements between the CPU and the GPU. This
goal required the ability to share resources like GPU memory or a CUDA
stream between modules, which is described in
Section~\ref{cudaTools:resource}. A mechanism to transfer data only
when necessary is then described in Section~\ref{cudaTools:transfer}.
The design of the tools should also be extendable to multiple non-CPU
device types, and be able to make use of multiple devices per type.

Much of the interaction with the CMSSW framework is done by a wrapper
template \cmsProductT{} for data products (of type \texttt{T}, which
itself can be partly or fully in the GPU memory), and a helper object
\cmsScopedContext{}\footnote{In reality the class has a couple of variants, but
  for the discussion in this paper grouping them into one is
  sufficient.} that is intended to be used in the body of the module's
\texttt{acquire()} and \texttt{produce()} functions.

The pattern has similar functionality as CUDA graphs~\cite{cuda}, that
is a directed acyclic graph of memory transfers, kernel launches, and
host functions, but at a higher level. In theory CUDA graphs could be
used in the background, but preliminary investigations indicate that
the current implementation of CUDA graphs is too restrictive for our
usage pattern.

\subsection{Asynchronous execution}
\label{cudaTools:async}

In order to avoid the CPU waiting for GPU work to finish only the
asynchronous CUDA runtime API calls may be used during event
processing. Essentially this constraint means memory transfers and
\texttt{memset} calls, because the kernel launches are asynchronous by
construction. The asynchronous API calls require the use of CUDA
streams. Work items queued in a single CUDA stream are
executed serially, but concurrently with respect to work in other CUDA
streams.

The \ProductT{} and \ScopedContext{} tools were developed such that
each parallel branch in the module DAG\footnote{The modules form a
  directed acyclic graph (DAG) by their data dependencies.}
automatically gets its own CUDA stream. With such an approach the
available concurrency is maximally expressed to the CUDA runtime,
which can then schedule work as it sees best.

In addition, possible synchronization points need to be carefully
avoided. These synchronization points include for example memory
allocations with the CUDA runtime API, explicit synchronization calls,
and calls to \texttt{assert()} in kernel code. The simplest way to
avoid dynamic memory allocations through the CUDA runtime API during
event processing would be to allocate the necessary device and pinned
host memory for each module at the beginning of the job. This approach
has, however, several drawbacks. Most importantly, it would lead to
allocation of much more memory than is actually needed at any given
time to cover all possible cases: 1) need to allocate memory for all
concurrent events even though not all of them will be processed by the
same module at the same time; 2) not all modules will be running
concurrently because of data dependencies; and 3) the allocated memory
would have to be large enough to cover the largest need of the
processed events, and typically there are large variations between
events.

To address all these drawbacks, we decided to use a memory pool for
both the device and the pinned host memory for the memory allocations
done on the host. At the time of writing these memory pools are based
on the \texttt{CachingDeviceAllocator} from the CUB
library~\cite{cub}. While this memory pool allocates its memory during
the event processing, essentially by caching the allocations, the cost
of the API calls gets amortized.

It should be noted that only modules that need to synchronize the GPU
and CPU for some CUDA stream, for example to transfer some data from
GPU to CPU, need to use the External Worker mechanism
(Section~\ref{externalWork}) instead of an explicit synchronization
call\footnote{See footnote 1}. Modules that only queue asynchronous
GPU work can call the CUDA runtime API directly, in a way that
resembles the \texttt{streaming\_node} in the TBB Flow Graph
library~\cite{tbb}.

The proper signaling of work completion is handled by the
\ScopedContext{}: the \WaitingTaskWithArenaHolder{} must be given to
the constructor of the \ScopedContext{}, and the destructor of
\ScopedContext{} queues a callback function into its CUDA stream with
\texttt{cudaStreamAddCallback()} to which the
\WaitingTaskWithArenaHolder{} is passed. The callback function then
notifies the \WaitingTaskWithArenaHolder{}, and in case of errors,
creates an exception object to be propagated.

\subsection{Sharing of resources between modules}
\label{cudaTools:resource}

A chain of modules with producer-consumer relationships on the data in
GPU memory run most efficiently if they agree at least on running
their work on the same device. Furthermore, it would be beneficial for
a linear chain of work to be queued into the same CUDA stream, and in
case of branches in the DAG, let the CUDA runtime to deal with the
synchronization between the branches.

The GPU data product wrapper \ProductT{} holds the device ID, the CUDA
stream where the producing work was queued into, and a CUDA event to
mark the completion of the asynchronous processing in case that did
not finish by the time the module's \texttt{produce()} function ended.
A module that queues more GPU work with \ProductT{} as an input
constructs \ScopedContext{} with the \ProductT{} as an argument. The
\ScopedContext{} then sets the current device based on the input
product, and re-uses the CUDA stream from the input product if the
module is the first one to ask it from the \ProductT{}. If another module
re-used the CUDA stream first, the \ScopedContext{} creates a new CUDA
stream and uses that.

The only way for a consumer to obtain \texttt{T} from \ProductT{} is
via the \ScopedContext{}. Upon request by the module, the
\ScopedContext{} checks whether the \ProductT{} uses the same CUDA
stream as the \ScopedContext{} was constructed with. If it does, the
\texttt{T} object can be returned immediately, because the sequential
nature of CUDA stream ensures the proper synchronization. In case the
CUDA streams are different, the availability of \texttt{T} is checked
via the CUDA event. If the CUDA event indicates that the asynchronous
work producing \texttt{T} has completed, the \texttt{T} can again be
returned safely. If the work is still incomplete, the \ScopedContext{}
introduces a wait on its CUDA stream on the CUDA event of the
\ProductT{} by calling \texttt{cudaStreamWaitEvent()} before returning
the \texttt{T} object.

\subsection{Minimization of data movements}
\label{cudaTools:transfer}

The CMSSW framework runs a producer module only if some other module
consumes the output data product of the producer module. We can make
use of this behavior to minimize the data transfers from GPU to CPU by
adding additional, specific modules that only queue the data transfers. This
way the transfers are avoided if no other module asks for a CPU copy of
data in GPU memory, but the capability to do the transfer exists in
case a module asking for such a copy is added into the configuration.

This approach works well together with the SwitchProducer mechanism in
the configuration (Section~\ref{switchProducer}), and the way data in
GPU memory is passed from one module to another
(Section~\ref{cudaTools:resource}). In practice the user is expected
to use the SwitchProducer to choose between the CPU module, and the
module that transfers the GPU data back to CPU, and leave all the
dependent GPU modules to be run by the framework.


\section{Summary}
\label{summary}

This paper described the generic building blocks we have developed for
CMSSW that can be used to continue the exploration of using non-CPU
resources for CMS data processing. We are exploring the performance
characteristics of the described pattern for using CUDA from the data
processing modules. An example of the achieved performance on a
real-world application can be found in~\cite{bocci:2020}.

\section*{Acknowledgements}

This document was prepared by the CMS Collaboration using the
resources of the Fermi National Accelerator Laboratory (Fermilab),
U.S. Department of Energy, Office of Science, HEP User Facility.
Fermilab is managed by Fermi Research Alliance, LLC (FRA), acting
under Contract No. DE-AC02-07CH11359.

\end{document}